\documentstyle[aps,preprint]{revtex}
\draft
\newcommand{\npb}{Nucl. Phys. B}
\newcommand{\be}{\begin{equation}}
\newcommand{\ee}{\end{equation}}
\newcommand{\ba}{\begin{eqnarray}}
\newcommand{\ea}{\end{eqnarray}}
\newcommand{\nn}{\nonumber}

\newcommand{\gm}{\gamma}
\newcommand{\szq}{s_{z}^q}
\newcommand{\kz}{k_{z}}
\newcommand{\kv}{k_{\perp}}
\begin{document}

%@twocolumnfalse\endcsname

%\thispagestyle{empty}

\title{
\hbox to\hsize{
\large  To appear in Phys. Rev. D Comments \hfil E-print: astro-ph/000000}
\vskip 1.0cm 
Comment on ``Can there be a quark-matter core in a magnetar?"
}

\author{
In-Saeng Suh$^\ast$, G. J. Mathews$^\dagger$, and F. Weber$^\ddagger$
}

\address{
Center for Astrophysics, Department of Physics, University of Notre Dame, \\
Notre Dame, Indiana 46556, USA \\
$^\ast$isuh@nd.edu, $^\dagger$gmathews@nd.edu, $^\ddagger$fweber@nd.edu
}

\date{\today}
\maketitle

\begin{abstract}
We comment on the paper by Tanusri Ghosh and Somenath Chakrabarty, 
Phys. Rev. D63, 043006 (2001).
In that paper it was argued that a first-order transition to quark matter  
in the core of a magnetar is absolutely forbidden
if its magnetic field strength exceeds $10^{15}$ G. However,
we show in this comment that if the quark anomalous magnetic moment 
and the population of higher Landau levels is taken into 
account, it may still be possible for a first-order phase transition from nuclear
matter to quark matter to occur in the core of a magnetar. 
These effects may also obscure the question of whether
beta equilibrium favors strange quark matter in the core of a magnetar.
%\keywords{stars: interiors --- stars: magnetic fields --- stars: neutron}
%\\
%\\
\end{abstract}

%\vskip 1.5pc]
%--------------------------------------------------
%\narrowtext
\newpage

Ghosh and Chakrabarty \cite{ghosh1,chakra} have recently investigated the
effects of a strong quantizing magnetic field on the nucleation of quark-matter 
droplets in the core of a magnetar. They argued that since the surface energy of the 
quark phase diverges logarithmically for the zeroth Landau level, 
a first-order phase transition from hadronic to quark matter would be
absolutely forbidden at the core of a magnetar. This is true as long as the
magnetic field strength exceeds $10^{15}$ G and only the zeroth Landau levels of 
light ($u,d$) quarks are populated. Here, however, we argue that 
if the quark anomalous magnetic moment and higher Landau levels are taken into
account, it is possible to avoid the divergence in the zeroth Landau level so that
a first-order phase transition from nuclear matter to quark matter may still be
possible in the core of a magnetar.

They also computed the chemical equilibrium and concluded that beta equilibrium
does not favor the existence of strange quark matter in the core of a magnetar. 
Here, we argue that, this conclusion may also be obscured by effects of the 
quark anomalous magnetic moment and higher Landau levels, though their conclusion 
probably remains valid.

The argument of Ghosh and Chakrabarty regarding the phase transition is as follows. 
The rate of nucleation of stable quark-matter 
droplets due to fluctuations in a metastable hadronic medium is given by \cite{landau}
\be
\Gamma = \Gamma_{0} e^{-{\sigma^3}/{C}} ,
\ee
where $\Gamma_{0}$ and $C$ are finite constants, and $\sigma$ is the surface
tension of the quark phase in metastable hadronic matter.

Then, using the MIT bag model of color confinement,  the expression
for the surface tension of the quark phase in the presence
of a strongly quantizing magnetic field of strength $B_m$ is given by \cite{ghosh1,chakra}
\ba
\sigma_i &=& \frac{2 T B_m}{8 \pi} Z_{i}^{q} \sum_{\nu=0}^{\infty} \int_{\nu=0}^{\infty} 
\frac{d \kz}{(\kz^{2} + \kv^{2})^{1/2}}  \nn \\
&~&\times {\rm ln}\Bigg(1 + exp \Bigg[- \frac{E_{\nu (i)} - \mu_i}{T} \Bigg] \Bigg) {\cal G},
\ea
where $i$ indicates the quark flavor ($u$, $d$, or $s$), $Z_{i}^{q}$ is the flavor
charge, $\nu$ is the Landau quantum number, $\kz$ and $\kv$ are the components
of momentum along and perpendicular to the $z$-axis,
and $\mu_i$ is the chemical potential.
The quark-energy eigenvalues are given by 
\be
E_{\nu (i)} = [m_{i}^{2} + \kz^2 + \kv^2 ]^{1/2}, 
\ee
with
\be
\kv {_{(i)}} = \sqrt{2 \nu Z_{i}^{q} B_m},
\ee
and the quantity ${\cal G}$ in Eq. (2) is defined as 
${\cal G} = 1 - 2{\rm tan^{-1}}(k/m_i) / \pi$.  

Now, through simple integration by parts, we can easily derive the surface tension 
of the quark phase and show that it diverges logarithmically for the lowest Landau 
level, $\nu = 0$. That is,  
the surface tension goes as $ \sim - {\rm ln}(\nu)$ as $\nu \rightarrow 0$
in the presence of a quantizing magnetic field.
There is also a factor $1/k = 1/(\kz^2 + \kv^2)^{1/2}$ in Eq. (2) which will diverge 
for $\kz = 0$ and $\nu=0$.
Therefore, the rate of nucleation of quark-matter droplets as given by Eq. (1) becomes
identically zero. It thus follows that if the magnetic field at the core
of a neutron star is strong enough to populate the Landau levels of quarks,
the nucleation of quark droplets becomes impossible, 
which means that under such conditions a first-order phase transition from hadronic 
matter into quark matter is absolutely forbidden. 
Ghosh and Chakrabarty also have insisted that if higher order as well as the zeroth  
order Landau levels are populated simultaneously, the conclusions do not change.

However, as mentioned in Ref. \cite{suh}, if ultra-strong magnetic fields can exist 
in the interior of neutron stars, such fields will affect the behavior of 
the residual charged particles. 
Moreover, contributions from the anomalous magnetic moment (AMM) of the particles
in a strong magnetic field should also be significant.
Therefore, we should include the AMM to derive a modified energy spectrum of the particles. 

Then, the energy dispersion relation for quarks $E_{q(i)}$ for an arbitrary Landau level
in a magnetic field becomes \cite{grandy}:
\be
E_{q(i)} = \kz^2 c^2 + \Bigg\{ m_i c^2 \Bigg[1 + 2 \gm_{i}^{q} n_{f}^{q(i)} \Bigg]^{1/2}
- \szq \mu_{f}^{i} B_m  Z_{i}^{q} \Bigg\}^2 ,
\ee
where $n_{f}^{q(i)} =  n + \frac{1}{2} - \szq Z_{i}^{q}$, $n$ is the  principal quantum
number of the Landau level, $\szq =\pm \frac{1}{2}$ is the $z$ component of the quark spin,
and $\mu_{f}^{i}$ is the quark magnetic moment discussed below. 
The summation over $\nu$ in Eq. (2) is now replaced with two summations over $n$ and $\szq$, 
i.e.,
\be
\sum_{\nu = 0}^{\infty} \Longrightarrow \sum_{n = 0}^{\infty}  \sum_{\szq}.
\ee

An estimate of the quark anomalous magnetic  moment (QAMM) can be obtained as follow.
In the Pauli notation for fermions with charge $e_f$, the electronmagnetic current up
to first order in the photon momentum $q_\nu$ is given by
\be
J^\mu = e_f \bar{u} \Gamma^\mu u = e_f \bar{u} 
\Bigg[F_{1} (q^2) \gamma^\mu 
+ F_{2} (q^2) \frac{i \sigma^{\mu \nu}}{2 m c} q_\nu \bigg] u, 
\ee
where the form factors $F_{1}(q^2)$ and $F_{2}(q^2)$ describe the electromagnetic 
structure of the charged particle. For free photons, $q^2 = 0$, and thus, 
the charge form factor $F_{1}(0) = 1$. 
The magnetic form factor $F_2 (0) \equiv \kappa_f$ stands for the anomalous part of 
the magnetic moment $\mu_f$. It is given by
\be
\mu_f = \mu_0 \kappa_f, \;\;\;\; {\rm with} \;\;\; \mu_0 = \frac{e_f \hbar}{2 m c},
\ee
and $m$ stands for the particle mass. 
For example, the magnetic moment of ground-state nucleons has the experimentally
measured values of $\kappa_p \simeq 2.79$ for the proton and $\kappa_n \simeq -1.91$ 
for the neutron.
Therefore, in the constituent quark model for light hadrons, we have
\be 
\mu_p = \frac{1}{3} (4 \mu_u - \mu_d), \;\;\; \mu_n = \frac{1}{3} (4 \mu_d - \mu_u),
\ee
which leads to a nonzero quark anomalous magnetic  moment of
\be
\mu_u = 1.852 \mu_N,  \;\;\;\; \mu_d = -0.972 \mu_N,
\ee 
with $\mu_N$, the nuclear magneton. 

There are several theoretical and experimental studies which indicate that 
quarks have a QAMM \cite{bicudo,singh,kopp,brekke,fajfer}. 
Recently Bicudo et al. \cite{bicudo} have shown that in the case of massless-current
quarks, chiral symmetry breaking usually triggers the generation of an anomalous
magnetic moment for the quarks. [Notice that the hadron $\leftrightarrow$ quark
phase transition is deeply related to chiral symmetry breaking.]
They also computed the QAMM in several quark models. 
Similarly, Singh \cite{singh} has also proven that, in theories in which chiral symmetry 
breaks dynamically, quarks can have a large anomalous magnetic moment.
In particular, K\"{o}pp et. al. \cite{kopp} have provided a stringent bound on 
the QAMM from high-precision measurements at LEP, SLC, and HERA.
To fit the measured magnetic moment of the baryon octet, it is found that the quarks 
must have a QAMM \cite{brekke}. Therefore, on the chiral symmetry breaking scale, 
if one considers the phase transition from hadronic matter to quark matter 
in a strong magnetic field, one should include the QAMM in the quark-energy spectrum.

Given that a QAMM likely exists, then our main point is that if instead of Eq. (3),
we use Eq. (5) which includes the QAMM,
we can at least avoid a divergence from the lowest Landau level, $\nu=0$, in calculations
of the surface tension [Eq. (2)] as well as in the curvature energy \cite{chakra} which
are included in the bubble nucleation rate.
Therefore, a finite value of the bubble nucleation rate can be obtained.
This means that it is still possible for a first-order phase transition to quark matter
to occur in a magnetar.

Even if a first-order phase transition to quark matter is possible,
however, one must still consider whether $uds$-quark matter
is the energetically preferred state in beta equilibrium.  
In \cite{ghosh1,ghosh2}, the chemical evolution of quark matter was studied. 
They concluded that in the presence of a strong magnetic field $\geq 4.4 \times 10^{13}$G, 
the matter in beta equilibrium is energetically unstable with respect to normal
neutron matter at the core. Hence, they concluded that the existence of
quark matter is impossible if the magnetic field exceeds the quantum
limit for electrons. An alternative way to understand this is that, since the quantum 
mechanical effect of a strong magnetic field on strange quarks is negligible 
unless the magnetic field strength is $\geq 10^{20}$G, it is energetically
favorable for the system to produce $d$-quarks whose Landau levels are populated,
or if the magnetic field is low enough so that only the electrons are affected,
then it is energetically much more favorable to produce electrons and $u$-quarks
(to balance the charge). As a result, the system in equilibrium is not $uds$-quark matter.

  Here we point out that their conclusions were based upon the presumption
that only the lowest Landau level was populated for all charged
particles (except the $s$-quark).  As shown in \cite{suh}, the Landau levels at high density
depend upon both the strength of the magnetic field and the Fermi energy.  
At high density, beta equilibrium ($i.e.$, charged particle fraction) is absolutely different 
between the case in which only the lowest Landau level is considered and 
the case in which higher Landau levels are included.
In particular, when the anomalous magnetic moment is included, one must consider
that the maximum landau level has a strong correction in the spin-polarization 
term (third term in Eq. (5)).  There is a good chance that
their conclusion may remain even if the ground states are recalculated
using the higher Landau levels.  Nevertheless, it is at least conceivable
that a revised calculation could shift the equilibrium back
to strange quark matter.  This calculation should be redone before we can 
categorically conclude that strange quark matter core is absolutely impossible for
a magnetar.

This work supported in part by DOE Nuclear Theory Grant DE-FG02-95ER40934.

%------------------------------------------------------------------------------------------

%-----------------------------------------------------------------------------------------

\begin{references}

\bibitem{ghosh1} T. Ghosh and S. Chakrabarty, \prd, {\bf 63}, 043006 (2001)

\bibitem{chakra} S. Chakrabarty, \prd ~51, 4591, (1995).

\bibitem{ghosh2} T. Ghosh and S. Chakrabarty, Int. J. Mod. Phys. D{\bf}10, 89 (2001)

\bibitem{landau} L. D. Landau and E. M. Lifshitz,
Statistical Mechanics (Clarendon: Oxford, 1938)

\bibitem{suh} In-Saeng Suh and G. J. Mathews,  2001,
\apj ~546, 1126; A. Broderick, M. Prakash, and J. M. Lattimer, 
\apj ~537, 351 (2000).  

\bibitem{grandy} W. T, Jr. Grandy,  
{Relativistic Quantum Mechanics of Leptons and Fields}, 
(Dordrecht: Kluwer Academic Publishers) (1991)

\bibitem{bicudo} P. J. de A. Bicudo, J. E. F. T. Ribeiro, and R. Fernandes,
\prc {\bf 59}, 1107 (1999).

\bibitem{singh} J. P. Singh, \prd, {\bf 31}, 1097 (1985).

\bibitem{kopp} G. K\"{o}pp, D. Schaile, M. Spira, and P. M. Zerwas,
Z. Phys. C {\bf 65}, 545 (1995).

\bibitem{brekke} L. Brekke, \npb {\bf 240}, 400 (1995).

\bibitem{fajfer} S. Fajfer and R. J. Oakes, \prd {\bf 28}, 2881 (1983).

\end{references}
\end{document}